\documentclass[pra,aps,showpacs,groupedaddress,superscriptaddress,twocolumn,toc=flat]{revtex4-1}

\usepackage[colorlinks=true,linkcolor=blue,urlcolor=blue,citecolor=blue]{hyperref}

\usepackage[utf8x]{inputenc}
\usepackage{color}
\usepackage{bbm} 

\usepackage{amsfonts,amsmath,amssymb,stmaryrd}

\usepackage{graphicx}
\usepackage{subfigure}  
\usepackage{bbm} 
\usepackage{hyperref}
\usepackage{epsfig}
\usepackage{mathrsfs}
\usepackage{verbatim}
\usepackage{centernot}
\usepackage{ulem}
\usepackage{nicefrac}

\usepackage{array}

\usepackage{cancel,ifthen}
\newcommand{\cmnt}[2][NoInPuT]{\ifthenelse{\equal{#1}{NoInPuT}}{}{{\color{red}\sout{#1}}} {\color{blue} #2}}

\usepackage{bm}	

\begin{document}
\normalem	

\title{Interpretable correlator Transformer for image-like quantum matter data}


\author{Abhinav Suresh}
\affiliation{Department of Physics, Indian Institute of Science Education and Research, Pune}
\affiliation{Institut für Theoretische Physik, Universität Regensburg, D-93035 Regensburg, Germany}

\author{Henning Schl\"omer}
\affiliation{Fakult\"at f\"ur Physik, Munich Quantum Center, Ludwig-Maximilians-Universit\"at M\"unchen}
\affiliation{Munich Center for Quantum Science and Technology (MCQST), Schellingstr. 4, D-80799 M\"unchen, Germany}
\author {Baran Hashemi}
\affiliation{ORIGINS Data Science Lab}
\affiliation{Technical University of Munich, M\"unchen, Germany}

\author{Annabelle Bohrdt}
\affiliation{Institut für Theoretische Physik, Universität Regensburg, D-93035 Regensburg, Germany}
\affiliation{Munich Center for Quantum Science and Technology (MCQST), Schellingstr. 4, D-80799 M\"unchen, Germany}

\date{\today}

\begin{abstract}
Due to their inherent capabilities of capturing non-local dependencies, Transformer neural networks have quickly been established as the paradigmatic architecture for large language models and image processing. Next to these traditional applications, machine learning methods have also been demonstrated to be versatile tools in the analysis of image-like data of quantum phases of matter, e.g. given snapshots of many-body wave functions obtained in ultracold atom experiments. While local correlation structures in image-like data of physical systems can reliably be detected, identifying phases of matter characterized by global, non-local structures with interpretable machine learning methods remains a challenge. Here, we introduce the correlator Transformer (CoTra), which classifies different phases of matter while at the same time yielding full interpretability in terms of physical correlation functions. The network's underlying structure is a tailored attention mechanism, which learns efficient ways to weigh local and non-local correlations for a successful classification. We demonstrate the versatility of the CoTra by detecting local order in the Heisenberg antiferromagnet, and show that local gauge constraints in one- and two-dimensional lattice gauge theories can be identified. Furthermore, we establish that the CoTra reliably detects non-local structures in images of correlated fermions in momentum space (Cooper pairs) and that it can distinguish percolating from non-percolating images.
\end{abstract}


\maketitle

\section{Introduction}
Machine learning (ML) techniques provide powerful tools for image analysis~\cite{Song1996, Ripley1996PatternRA}. Since the dawn of deep neural networks, considerable progress has been made in image classification (computer vision)~\cite{Krizhevsky2012, Donahue2014, Rowley1998,dosovitskiy2021image}. In recent years, this has led to an increasing interest in applying machine learning based analysis tools to a wide range of physically relevant input data~\cite{Carleo2019}. For instance, it has been demonstrated that neural networks can efficiently classify and differentiate between various phases of matter~\cite{wang2016, Carrasquilla_2017,van_Nieuwenburg_2017}, including topologically ordered states~\cite{Zhang_2018, sun2018, Huembeli2018, Yoshioka2018, Carvalho2018, Rodriguez_Nieva_2019, Holanda2020} and systems with strong correlations~\cite{Ch'ng2017, Broecker_2017}. Furthermore, progress has been made towards discovering novel quantum phases~\cite{Hu2017, Miles2023, Cassella2023}; further unsupervised and supervised learning of quantum phases and their transitions have been studied in Refs.~\cite{Ch'ng2018,Wetzel2017, Tanaka_2017, Zhang2017, Beach2018, Wang2017, broecker2017quantum, Suchsland2018,  Iakovlev2018, Schafer2019,Lozano-Gomez2022, Kottmann_2020, SALCEDOGALLO2020166482}. 

\begin{figure*}
    \centering
    \includegraphics[width=\textwidth]{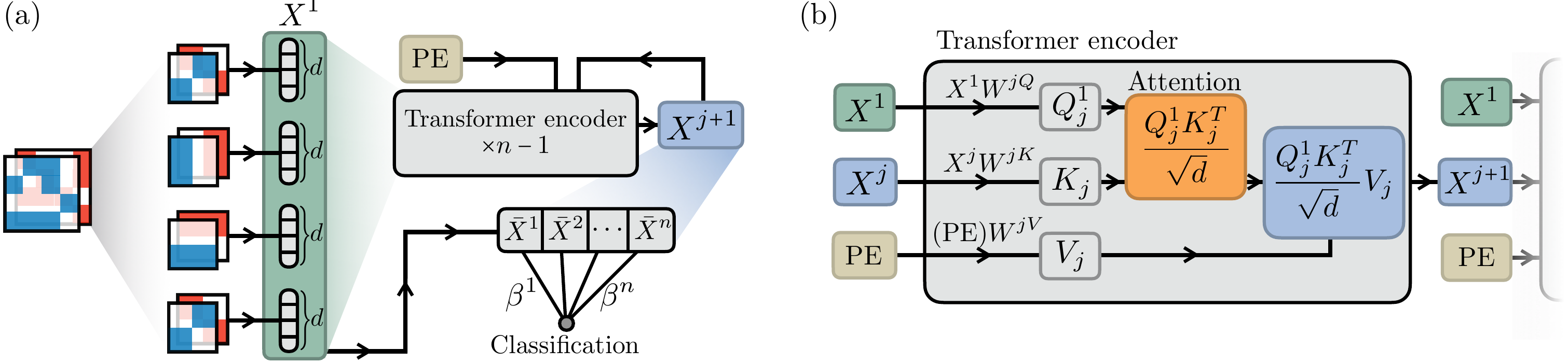}
    \caption{\textbf{The correlator Transformer.} (a) The overall flow starts from the 2D input data, here illustrated by a snapshot with two channels. After cutting the input into patches, they are linearly embedded to dimension $d$. The collection of $d$-dimensional vector tokens is denoted by $X^1$ (green box). Higher-order correlations are then generated in the Transformer encoder. To analyze $n^{\text{th}}$ order correlations, $n-1$ encoding steps are carried out. While the query remains identical throughout all layers, keys ($K_j$) and values ($V_j$) are learned and computed in each layer, shown in panel (b). In the $j^{\text{th}}$ layer, the attention then includes correlations up to order $j+1$. By multiplying with the value $V_j$, $X^{j+1}$ is computed. PE denotes the absolute positional encoding. Averaging the outputs results in the vector $[\bar{X}^1, \bar{X}^2,\dots,\bar{X}^n]$, which is then used for classification. Through a regularization path analysis, weights $\beta^{\ell}$, $\ell = 1,\dots,n$ of the final layer give information about which correlations are important for successful classification.}
    \label{fig:corr_vit}
\end{figure*}

With the development of highly controllable analog quantum simulation platforms, e.g. ultracold atoms in optical lattices~\cite{Bloch2008, Esslinger2010, Bohrdt2020}, the application of neural networks to projective measurements of the many-body wave function constitutes one particular field of interest~\cite{Bohrdt_ML}. These snapshots of the wave function contain a plethora of information that goes beyond local observables~\cite{Islam2015, Hilker2017, Sompet2022}, making them a promising platform for machine learning tools~\cite{Kamig2021, Schloemer2022_recon,Miles2023}.

In order to gain physical insights, interpretable machine learning techniques that are able to detect underlying order in a given set of measurements are desirable. 
Examples for physically interpretable algorithms include finding order parameters using support vector machines (SVM)  \cite{Ponte2017, Liu2019, Greitemann2019}, interpretability in terms of principle component analysis (PCA) \cite{Wetzel_2017, Costa2017,Hu2017} and analyzing the Hessian matrix of the loss function~\cite{Dawid_2021}. 
Using network theory, wave function networks have been developed to study such quantum simulation snapshots \cite{Mendes_Santos_2024}. 
In relatively shallow neural networks, insights can potentially be gained by looking into learned weights for a given classification task~\cite{Suchsland2018, Bl_cher_2020,Khatami_2020}. 

Typically, however, despite their effectiveness, neural networks tend to operate as black boxes. 
\emph{Interpretable} neural network architectures ideally combine a high performance with the possibility to uncover patterns and correlations in physical systems with complex dependencies. 
For this purpose, the correlation convolutional neural networks (CCNN)~\cite{Miles2020CorrelatorCN, Henning2023, Miles2023} has been developed, which allows to extract the most defining correlations in the Fock basis of a given snapshot. However, the convolutional structure of the CCNN architecture prevents an efficient detection of global structures and hidden or non-local order.

In order to access long-range dependencies in a snapshot, a different choice of architecture is more natural:
The introduction of the Transformer architecture \cite{vaswani2023attention} marked a groundbreaking moment in the development of large language models. Although conceived as a sequence-to-sequence model, this was later adapted to the vision Transformer \cite{dosovitskiy2021image} to be compatible with image data. The attention mechanism, which aims to mimic human-like attention, provides a map of pertinent portions and correlations of the image, which may help to gain a physical understanding of (global) underlying structures. It has been shown that the Transformer architecture can be used to learn for example entanglement growth and state complexities from experimental data \cite{kim2024attentionquantumcomplexity}. In~\cite{Henning2023}, the Transformer architecture was applied to quantum snapshots of the Heisenberg model, however, without giving a physical interpretation of the resulting attention mechanism. Here, we demonstrate that the vision Transformer can reliably classify images of correlated systems, with full interpretability in terms of local and non-local correlation functions. In particular, we introduce a tailored attention mechanism that enables us to isolate and capture specific orders of correlations from different Transformer layers.

This paper is organized as follows. In Section~\ref{Corr_vit}, we introduce the correlator Transformer architecture. We discuss how higher-order correlators are constructed in each subsequent layer and demonstrate that by adding tailored $L_1$ regularization, one can obtain the most relevant order of correlation. After obtaining the latter, an analysis of the learned weights of that particular Transformer layer (attention maps) enables further insights into the specific correlator. In Section~\ref{local_obs}, we show our results for local correlations upon applying the correlator Transformer to different physical models: we demonstrate how the Transformer learns local correlations in the two-dimensional (2D) antiferromagnetic (AFM) Heisenberg model, the one-dimensional (1D) $\mathbb{Z}_2$ lattice gauge theory (LGT) as well as the 2D $\mathbb{Z}_2$ Ising LGT, using local correlations up to fourth order. In Section~\ref{nonlocal_obs}, we apply the Transformer to problems involving non-local correlations, in particular in the analysis of momentum space images of Cooper pairs and in the classification of percolating and non-percolating snapshots. Finally, we summarize our results in Section~\ref{outlook}.

\section{Correlator Transformer}\label{Corr_vit}
The correlator Transformer (CoTra) architecture is inspired by the vision Transformer, a state-of-the-art image classification model. Like the vision Transformer, the CoTra is a versatile model that can handle different types of images. In this work, the input to our model is image-like quantum matter data, including, for instance, projective measurements in real or momentum space of quantum gas microscopes~\cite{Bloch2008, Esslinger2010} or tunneling microscope images~\cite{Fujita2014}.

For a given depth $n$ of the CoTra (corresponding to the number of layers in the Transformer architecture), the model has access to $n+1$ order weighted correlations (from now on denoted by $X^{n+1}$) with respect to these input measurements. These weights then give insights into the specific correlations that the model uses for its classification.

\subsection{Architecture}
Consider a snapshot in $\mathbb{R}^{(h,h,c)}$ where $h,c$ are the height (width) and number of channels. As done in standard vision Transformer workflows, we first divide this snapshot into equal patches of dimension $p\times p$, where $p$ is the patch size; hence, the transformed snapshot is of dimension $(h/p)^2 \times cp^2$, where $M = (h/p)^2$ denotes the number of patches. Afterwards, a linear projection is performed that transforms the patches to vectors of hidden dimension $d$. We label this projected snapshot as $X^1$ $\in$  $\mathbb{R}^{((h/p)^2,d)}$; it can be directly used as input of the Transformer encoder or it can be supplemented with additional learnable positional encoding depending on the input data, as further commented on below. As the above only consists of reordering and linear transforming the original data, the resulting matrix corresponds to first order terms $X^1$, see Fig.~\ref{fig:corr_vit}~(a). By default, we do not use any $Layer~ Norm$, and it is used only if it is crucial for the model's performance. This choice is made to keep the model as simple as possible for better interpretability.

The general idea of the CoTra is as follows (see Fig.~\ref{fig:corr_vit}): In the first layer, both the query and key projections are computed from $X^1$, which, together with a positional embedding, returns the second-order terms $X^2$. In a second Transformer layer, the query projection remains the same, while the key projection is now computed from the output of the first layer, i.e. $X^2$. This second layer now computes the third-order terms $X^3$ and so on. Therefore, using $n$ Transformer layers enables us to compute correlations up to order $(n+1)$. 

In particular, the query and key for the first Transformer layer can be written as 
\begin{equation}
    \begin{aligned}
            Q_1 &= X^1 W^{1Q} \\
            K_1 &= X^1 W^{1K}, 
    \end{aligned}
\end{equation}
Where $W^{1Q}, W^{1K}$ are the query and key projection matrices of the first layer. Meanwhile, the value is a linear projection of the sine-cosine encoding, defined by~\cite{vaswani2023attention}
\begin{equation}
\begin{aligned}
\text{PE}_{(i, 2j)} &= \sin\left(\frac{i}{100000^{2j/d}}\right) \\ 
\text{PE}_{(i, 2j+1)} &= \cos\left(\frac{i}{100000^{2j/d}}\right).
\label{eq:sincos}
\end{aligned}
\end{equation}

Here, \( i \in \{0,1,...N-1\}\) is the position of $N$ patches and \( j \in \{0,1,....,d-1\} \) where $d$ is the hidden dimension. 
 Since these are not learnable parameters, they do not increase the complexity of the model. The value, then, is
 \begin{align}
     V_1 = (\text{PE}) W^{1V}.
 \end{align}

The Transformer output of the first layer ($X^2$) can now be written as [see Fig.~\ref{fig:corr_vit}~(b)]
\begin{equation}
\begin{aligned}
    X^2 &=\frac{1}{\sqrt{d}} Q_1 K^T_1V_1 \\
    &= \frac{1}{\sqrt{d}}X^1 W^{1Q} W^{1K^T}X^{1^T} (\text{PE}) W^{1V}.
\end{aligned}
\end{equation}

In particular, $X^2$ and the corresponding attention map $a^2 = \frac{1}{\sqrt{d}}Q^1K^{1^T}$ both contain only $2^{\text{nd}}$ order terms. This can be directly seen by writing out the matrix elements [see Eq.~\eqref{eq:a2A} in Apendix~\ref{sec:C2}] of the $2^{\text{nd}}$ order attention map (with patch labels $i,j$),  
\begin{equation}
\begin{aligned}
    a^2_{ij}  = \sum_{rs=1}^{cp^2}\sum_{klsm=1}^{d} x_{ir}x^T_{sj}P_{ik}P^T_{lj}&W^P_{rk}\\&W^{P^T}_{ls}W^{1Q}_{km}W^{1K^T}_{ml}.
    \label{eqn: attn2}
\end{aligned}
\end{equation}
Here, $x$ denotes the pixel values (projective spin measurements, phase space measurements etc.) of a given snapshot,  $W^P$ is the linear projection matrix to the hidden dimension, $W^{1Q}$, $W^{1K}$ are the query and key projection matrices of the first layer and $P$ is an additional learnable positional encoding. Depending on the dataset, this learnable positional encoding is included at $X^1$ along with the sine-cosine encoding to improve accuracy. 
As the indices in Eq.~\eqref{eqn: attn2} $ i,j$ go through all the patches and $r,s$  through all the pixels inside the patches, we can obtain all the $2^{\text{nd}}$ order local and non-local correlators and their corresponding attention weights. 

Similarly, the $n^{\text{th}}$ layer output [see Fig.~\ref{fig:corr_vit}~(b)] is defined as
\begin{align}
     X^{n+1} = \frac{1}{\sqrt{d}}Q^{1}_n K^T_{n} V_{n}, 
\end{align}
where $Q^{1}_n$ denotes the query projection of $X^1$ by $W^{nQ}$, $K_{n}$ is the key projection of $X^{n}$ and $V_{n}$ is the value projection of the positional embedding at each layer $n$. This attention can be thought of as weighing the different correlations in $QK^T$ with their positional information. In this manner, we can build up further orders of weighted correlations using the Transformer, and by inspecting the weights of the corresponding attention maps, the most relevant correlators can be obtained. 

For each order $X^n$, we compute the mean $\bar{X}^n$ over all dimensions, which is used as input for the final classification, $\mathbf{\tilde{x}_n} = [\bar{X}^1,\bar{X}^2, \dots, \bar{X}^n]$. A linear transformation on $\mathbf{\tilde{x}_n}$ is then performed to form the logits for the cross-entropy loss function coupled with a regularization loss of strength $\alpha$. In particular, the loss function reads
\begin{align}
    \mathcal{L} \equiv -\sum_{i=1}^{N} y_i \log(\hat{y}_i) + \alpha \|\mathbf{\tilde{x}_n}\|,
\end{align}
where $y$ is the true distribution, $\hat{y}$ is the predicted distribution (logits), and $||.||$ is the norm.

This modified architecture contains a similar number of parameters as a traditional vision Transformer and is able to provide direct interpretability in terms of local as well as non-local correlations.
After training the CoTra with a particular hidden dimension ($d$), learning rate ($lr$), regularization strength ($\alpha$) and other hyperparameters, we collect the learned correlators and perform a regularization path analysis to understand which order terms influence the classification the most, as described next.
 
\subsection{Regularization Path Analysis}
Regularization path analysis is a common technique used for feature selection \cite{Tibshirani1996RegressionSA}. Here, as done in~\cite{Miles2020CorrelatorCN}, we use the $L_1$ regularization. The idea is to collect the learned mean correlators $\bar{X}^n$ and perform a lasso regression to see which of the $n$-point correlators are the most important for phase classification. We only re-train the architecture's logistic classifier part (final linear projection) rather than the whole model.

\begin{figure*}
    \centering
\includegraphics[width=\textwidth]{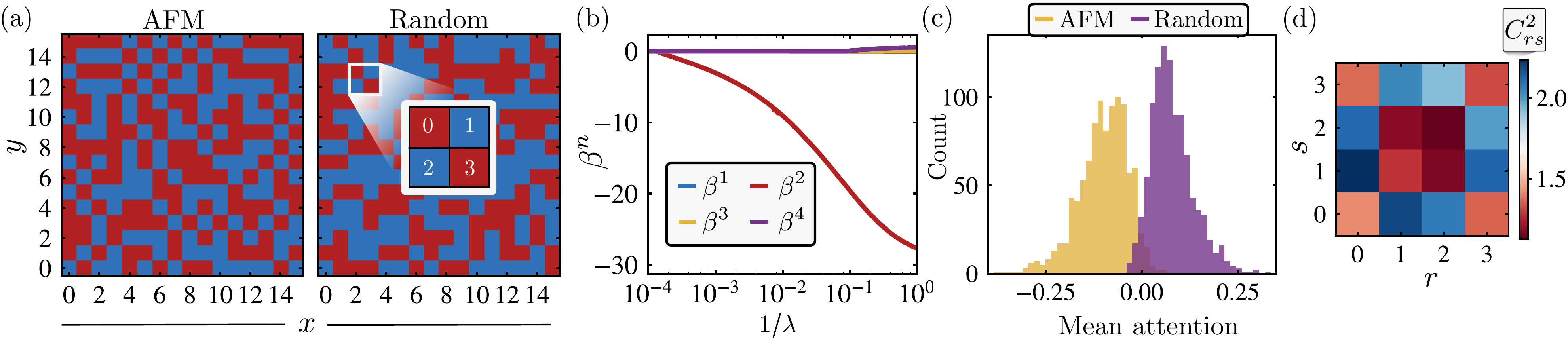}
    \caption{\textbf{2D Heisenberg model.} (a) Exemplary Heisenberg AFM (left) and random (right) configuration. The inset illustrates the labeling of the spins inside a $2\times2$ patch. (b) Regularization path analysis showing that $2^{\text{nd}}$ order correlations are used for classification. (c) Mean of the 2nd order attention map for the two data sets. The two distributions are almost entirely separate, underlining that the network uses second-order correlations for classification. (d) Intra-patch correlation weights $C_{rs}^2$, whereby nearest neighbor correlations show the most prominent signals.}
    \label{fig:H_afm_snapshots}
\end{figure*}

The loss function for the regularization path analysis is given by
\begin{align}
    \mathcal{L}_{\text{path}}\equiv -\sum_{i} y_i \log(\hat{y}_i) + \lambda \|\mathbf{\beta}\|,
    \label{L_path}
\end{align}
where $\beta \equiv [\beta^1,\beta^2,\dots, \beta^n]$ are the logistic weights of the corresponding (mean) $n$-point correlator. 
We use the Scikit-learn \cite{scikit-learn} logistic regression library to optimize the classifier over a range of $\lambda$ starting from a very high value, where the classification fails, to gradually smaller strengths where the classification is successful (see also~\cite{Miles2020CorrelatorCN}). By looking at the strengths of $\beta^n$ over the range of $\lambda$ one can see which particular $\beta^n$ (and hence which order of correlation) is responsible for a successful classification. Afterward, a closer look into the attention weights inside that particular Transformer layer tells us which specific $n$-point correlators are used by the network for classification. For local correlations, we can fix a patch and look at the attention weights inside that patch to see how each correlator is weighed. On the other hand, the attention map itself gives information about the dependencies between patches for long-range correlations.

\section{Learning Local correlations and gauge constraints}\label{local_obs}
In this section, we discuss the results of our model in learning local correlations. In the following, we consider three exemplary datasets: the 2D Heisenberg model, a 1D $\mathbb{Z}_2$ lattice gauge theory (LGT) and the 2D $\mathbb{Z}_2$ Ising LGT.

\subsection{2D Heisenberg model}
We start by analyzing snapshots of a prototypical spin model, namely the Heisenberg model. The Hamiltonian of the system is given by
\begin{align}
    \hat{H} = J \sum_{\langle i,j \rangle} \left( \hat{S}_i^x \hat{S}_j^x + \hat{S}_i^y \hat{S}_j^y + \hat{S}_i^z \hat{S}_j^z \right),
\end{align}
where $J$ is the nearest neighbour coupling constant and $S^{x,y,z}$ are spin-$1/2$ operators. We train the Transformer on snapshots from an equilibrium finite temperature state at $T/J = 0.6$ with AFM couplings ($J>0$) against random snapshots with vanishing magnetization. This ensures that the model does not compare magnetization values directly, but instead looks for higher-order correlations when classifying the snapshots.
We use a lattice size of $16 \times16$ and sample Heisenberg snapshots  using stochastic series expansion quantum Monte Carlo \cite{QGasML,Sandvik1991,Sandvik1999}. Fig.~\ref{fig:H_afm_snapshots}~(a) depicts one of the Heisenberg snapshots as well as a random configuration. The chosen hyperparameters for this data are $d=16, \text{patch size} =2,\lambda=10^{-3}$, and $lr=5\times10^{-3}$. The inset in the right panel of Fig.~\ref{fig:H_afm_snapshots}~(a) shows how the spins inside a single $2\times 2$ patch are numbered, which is important when evaluating the correlations at a later stage. We obtain an accuracy of 96.4\%.

\begin{figure*}
    \centering
    \includegraphics[width=0.97\textwidth]{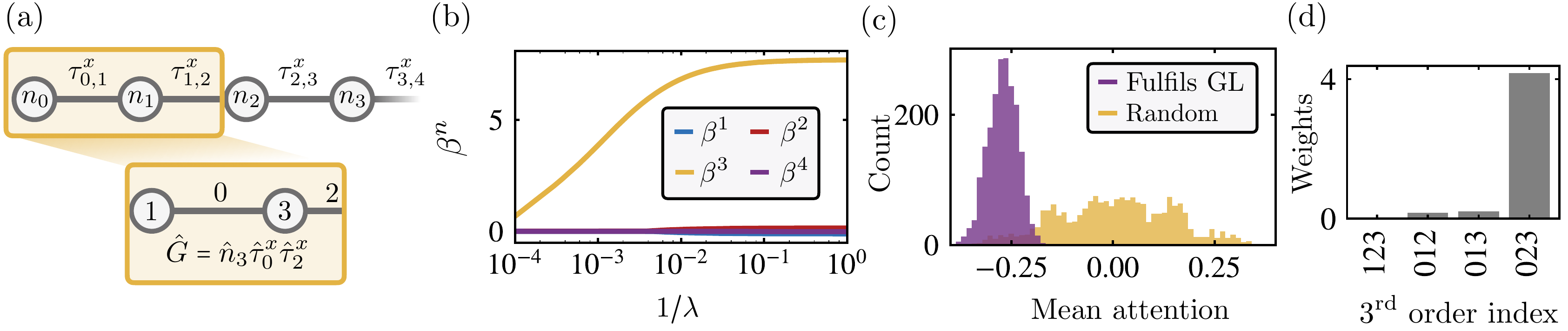}
    \caption{\textbf{1D lattice gauge theory.} (a) 1D chain indicating the matter sites $n_j$ and link sites $\tau^x_{j,j+1}$. A patch size of 2 is chosen (which contains two matter and two link sites), indicated by the yellow box. The numbering of elements inside the patch is shown in the inset. (b) Regularization path analysis showing $3^{rd}$ order correlations to be crucial for classification. (c) Mean of the $3^{rd}$ order attention map for the two states. While random snapshots show a broad distribution, the data that fulfils the gauge constraint features a sharp peak. (d) Third-order intra-patch correlation weights. Large weights attributed to correlators between the elements $0,2,3$ indicate that the CoTra is indeed checking if the gauge constraint is fulfilled.}
    \label{fig:1dlgt_results}
\end{figure*}

Fig.~\ref{fig:H_afm_snapshots}~(b) shows the regularization path, which indicates that the model is capable of classifying the two datasets by evaluating $2$-point correlations. We take the mean over the $2^{\text{nd}}$ order attention map $a^2 = \frac{1}{\sqrt{d}}Q^1K^{1^T}$ for each snapshot in the dataset and plot a histogram. Indeed, Fig.~\ref{fig:H_afm_snapshots}~(c) shows that the distributions of the mean $2^{\text{nd}}$ order attention for the AFM and random snapshots separate to a large degree, corroborating that the network uses $2^{\text{nd}}$ order correlations for its classification. Now, to see which particular $2^{\text{nd}}$ order correlators are used by the network, we take a more detailed look into the attention weights. The learned weights of the $2^{\text{nd}}$ order correlator (labels inside a patch are given by $r,s$) can be written as
\begin{equation}
\begin{aligned}
    C^{^{(2)}}_{rs} =\Bigg|\frac{1}{M} \sum_{i=1}^{M} \sum_{klm=1}^{d} P_{ik}&P_{li}^TW^P_{rk} \\&W^{P^T}_{ls}W^{1Q}_{km}W^{1K^T}_{ml}\Bigg|,
    \label{eq:C2}
\end{aligned}
\end{equation}
where we have taken the average over all patches $i$. $P$ is the learnable positional encoding, $W^P$ is the linear projection matrix to the hidden dimension and $W^{nQ}, W^{nK}$ denotes the query, key projections in the $n^{\text{th}}$ Transformer layer, see also Appendix~\ref{sec:C2} for an explicit derivation of Eq.~\eqref{eq:C2}.

Fig.~\ref{fig:H_afm_snapshots}~(d) depicts this $2$-point correlation matrix where $r,s$ runs through all four spins inside the patch. We can see that the model puts its emphasis on nearest neighbour correlations, cf. the labeling of spins within the patch in the inset of Fig.~\ref{fig:H_afm_snapshots}~(a). With the combined information of the regularization path and the learned weights, we conclude that the correlator Transformer is calculating second-order (intra-patch) correlations to make its classification decision. 

\subsection{1D $\mathbb{Z}_2$ LGT}
Originally proposed in the context of non-perturbative theories in particle physics, lattice gauge theories have become a paradigmatic class of systems also in the field of strongly correlated phases of matter~\cite{Levin2005, Lee2006}. Their defining feature is the presence of a local gauge symmetry.  Here, we consider a (1+1)D $\mathbb{Z}_2$ LGT where hard-core bosonic matter is coupled to $\mathbb{Z}_2$ gauge fields. The Hamiltonian of the system is described as~\cite{Zohar2017, Borla2020, Kebric2021, Halimeh2022} 
\begin{align}
\hat{H}_0 = J \sum_{j=1}^{L-1} \left( \hat{a}^\dagger_j \hat{\tau}^z_{j,j+1} \hat{a}_{j+1} + \text{H.c.} \right) - h \sum_{j=1}^{L} \hat{\tau}^x_{j,j+1},
\end{align}
where $ \hat{a}^\dagger_j,\hat{a}_{j}$ are the bosonic ladder operators acting on matter site $j$, $\tau^z_{j,j+1}$ denotes the gauge field between the matter site $j$ and $j+1$ and  $\tau^x_{j,j+1}$ is the electric field between the matter site $j$ and $j+1$. 
The $\mathbb{Z}_2$ electric field is subject to Gauss' law (GL), defined by the local generator 
\begin{align}
 \hat{G}_j = (-1)^{\hat{n}_j} \hat{\tau}^x_{j-1,j} \hat{\tau}^x_{j,j+1},
\end{align}
which has eigenvalues $g_j=\pm 1$. In the following, we test whether the CoTra can identify the local gauge structure. In order to do so, we generate two classes of snapshots: One set consists of states with $g_j=+1$ $\forall j$ in the lattice. The network is then given the task to distinguish this set from snapshots with random gauge sectors $g_j = \pm 1$ $\forall j$.

We consider a chain of length $16$ with open boundaries. Fig.~\ref{fig:1dlgt_results}~(a) depicts the left end of the 1D chain with matter sites $n_j$ and link variables $\tau^x_{j,j-1}$. In the following, a single patch consists of two matter and link variables, with labelling as depicted in the inset of Fig.~\ref{fig:1dlgt_results}~(a). For better interpretability, we describe both matter and gauge fields on the same footing, i.e., we convert the occupancy of the matter site (which is measured in the Fock basis as 0 or 1) to $\pm 1$. We note that this transformation changes the local generator of gauge invariance to 
\begin{align}
    \hat{G}_j = \hat{n}_j \hat{\tau}^x_{j-1,j} \hat{\tau}^x_{j,j+1}.
 \label{gauge_inv}
\end{align}
For the matter site corresponding to $n_1$ within the unit cell depicted in Fig.~\ref{fig:1dlgt_results}~(a), only the elements $0,2,3$ contribute to the calculation of Eq.~\eqref{gauge_inv}.

\begin{figure*}
    \centering
    \includegraphics[width=0.82\textwidth]{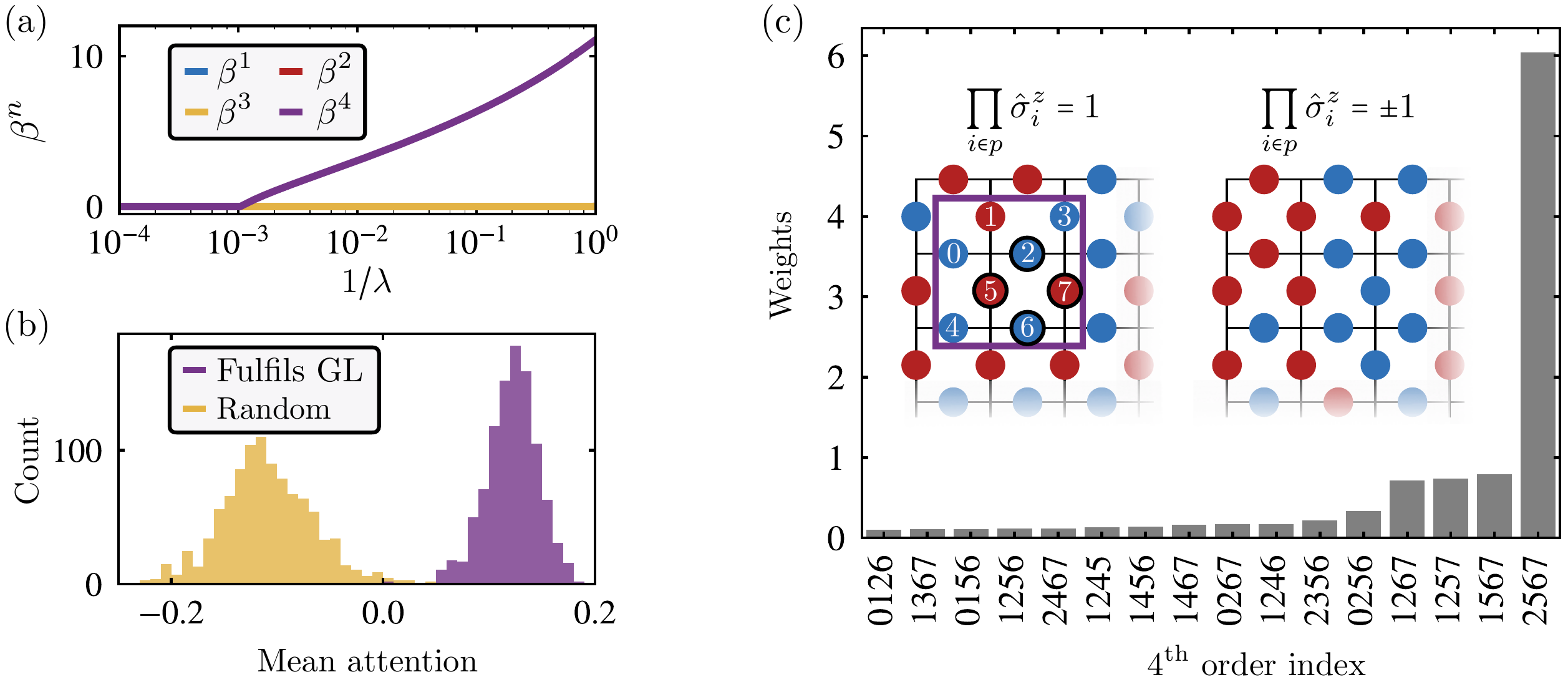}
    \caption{\textbf{2D Ising lattice gauge theory.} (a) After training, a regularization path analysis shows that $4^{\text{th}}$ order correlations are used to classify snapshots that fulfil Gauss' law from ones that do not. Indeed, the mean ($4^{\text{th}}$ order) attention values fully separate the two data sets, shown in (b). (c) Inset: Upper left part of snapshot fulfilling (not fulfilling) Gauss' law Eq.~\eqref{eq:GL2d} on the left-hand (right-hand) side. The labeling within each unit cell (purple box) is shown. The fourth order term on the plaquette that checks Gauss' law in each unit cell is given by $\hat{\sigma}^z_2 \hat{\sigma}^z_5 \hat{\sigma}^z_6 \hat{\sigma}^z_7$, illustrated by the black circles. Indeed, weights of $4^{\text{th}}$ order intra-patch correlations show significant weight for indices 2567, underlining that the CoTra learns to check Gauss' law. For illustrative reasons, only weights above a threshold of 0.1 are shown in the main panel of (c).}
    \label{fig:ilgt_results}
\end{figure*}

The hyperparameters when training the CoTra for this data are $d=12, \text{ patch size} =2~(\text{contains 2 matter and 2 links}), \lambda=2\times10^{-4}, lr=10^{-2}$ and we obtain an accuracy of 97\%. After training, we again perform a regularization path analysis, see Fig.~\ref{fig:1dlgt_results}~(b). Indeed, third-order correlations are used to distinguish the two different classes as expected from Eq.~\eqref{gauge_inv}. When evaluating the mean third-order attention maps for each snapshot, we see that states that fulfil the constraint $g_j = 1$ form a sharp peak, while random snapshots have a broad distribution [Fig.~\ref{fig:1dlgt_results}~(c)]. Indeed, the two histograms only have marginal overlap, underlining that the network is making its decision based on three-point correlations.

In order to analyze which particular three-point correlator within a patch is used for classification, we construct the third-order attention weights $a^3 = \frac{1}{\sqrt{d}}Q^{1}K^{2^K}$. The weights for the different possible three-point correlations are given by (see Appendix~\ref{sec:C3})
\begin{equation}
\begin{aligned}
        C^{^{(3)}}_{ers} = &\bigg|\frac{1}{M} \sum_{i=1}^M\sum_{kln=1}^{d} \sum_{abcm=1}^{d}P_{ia}P^T_{ki}P_{il}W^P_{ea}W^{P^T}_{kr}\\ &W^P_{sl}W^{1Q^T}_{mk}W^{2Q}_{ac}W^{1K}_{lm}W^{2K^T}_{cb}W^{1V^T}_{bn}(\text{PE})_{ni}\bigg|,
        \label{eq:weights3}
\end{aligned}
\end{equation}
where we again average over all patch labels $i$ and the indices $e,r,s$ go through the $4$ elements inside each patch. We note that in contrast to Eq.~\eqref{eq:C2}, the third-order correlator weight matrix explicitly includes the absolute positional embedding, as it involves terms ($X^2$) that have passed through the first Transformer layer. 

Fig.~\ref{fig:1dlgt_results}~(d) shows that significant weight is only on the correlator with indices $(e,r,s)=(0,2,3)$, which exactly corresponds to evaluating Eq.~\eqref{gauge_inv}. We note that choosing a larger patch size leads to identical conclusions. For instance, for a patch size of four, there are three possible combinations that contribute to the Gauss law; correspondingly, this leads to the appearance of three peaks in the correlation weights, i.e., the network calculates identical correlations to perform its classification. We were also able to classify states with 1-2 Gauss's law violations along the chain against the completely fulfilling snapshots with the addition of $Layer~Norm$ for the query and key projections. 

\subsection{Classical 2D $\mathbb{Z}_2$ Ising LGT}

Lastly, we focus on the classical $\mathbb{Z}_2$  Ising lattice gauge theory in 2D, where Gauss law is given by a $4^{\text{th}}$ order term. The Hamiltonian is defined as \cite{Greplova_2020}
\begin{align}
\hat{H} = -J \sum_p \prod_{l \in p} \hat{\sigma}^z_l,
\end{align}
where $J$ denotes the magnetic coupling constant, $\hat{\sigma}^z$ is the $z$ Pauli matrix, and the summation covers all plaquettes $p$ in the lattice. In the ground state, each plaquette $p$ of the lattice satisfies the Gauss' law:
\begin{align}
G_p = \prod_{l \in p}\hat{\sigma}^z_l = 1
\label{eq:GL2d}
\end{align}
The system has a highly degenerate ground state, comprising all possible configurations where $G_p = 1$ is obeyed.  At finite temperature, the gauge constraint is violated locally. Excited states are thus characterized by 
\begin{align}
G_p = \prod_{l \in p}\hat{\sigma}^z_l = \pm1.
\label{Gauss's_law}
\end{align}
As in the 1D LGT, we generate two classes of snapshots: One where Gauss's law ($G_p = +1$) is obeyed globally and one where it is violated randomly.

In particular, we consider a $16\times16$ lattice, with each lattice point having one horizontal and vertical link. The left-hand side of the inset in Fig.~\ref{fig:ilgt_results}~(c) depicts an exemplary configuration of the ground state, where on each plaquette, the product of Pauli $z$ operators is positive. On the other hand, the products are both positive and negative in a finite temperature snapshot; see the right-hand side of the inset in Fig.~\ref{fig:ilgt_results}~(c). It is important to note that a single patch of patch size $2$ contains $8$ links. We arrange the input snapshots into $2$ channels, one for the horizontal and one for the vertical links. By construction, out of the $8$ links in a patch, only the links $2,5,6,7$ contribute to the central plaquette of the patch; see the explicit labeling in the inset of Fig.~\ref{fig:ilgt_results}~(c).

The hyperparameters for this data are $d=6, \text{patch size} =2, \lambda=2\times10^{-4}, lr=5\times10^{-3}$. For these parameters, the network is able to fully classify the snapshots with an accuracy of $100\%$.
A regularization path analysis for the given choice of hyperparameters shows that the model is focusing on fourth-order correlators, Fig.~\ref{fig:ilgt_results}~(a). Note that we do not employ any additional learnable positional encoding here because the model is able to achieve perfect accuracy with absolute positional encoding. This is corroborated by the distributions of mean attention scores, as shown in Fig.~\ref{fig:ilgt_results}~(b). Indeed, both distributions fully separate, making a full distinction between the two data sets possible. 

\begin{figure*}
    \centering
    \includegraphics[width=0.9\textwidth]{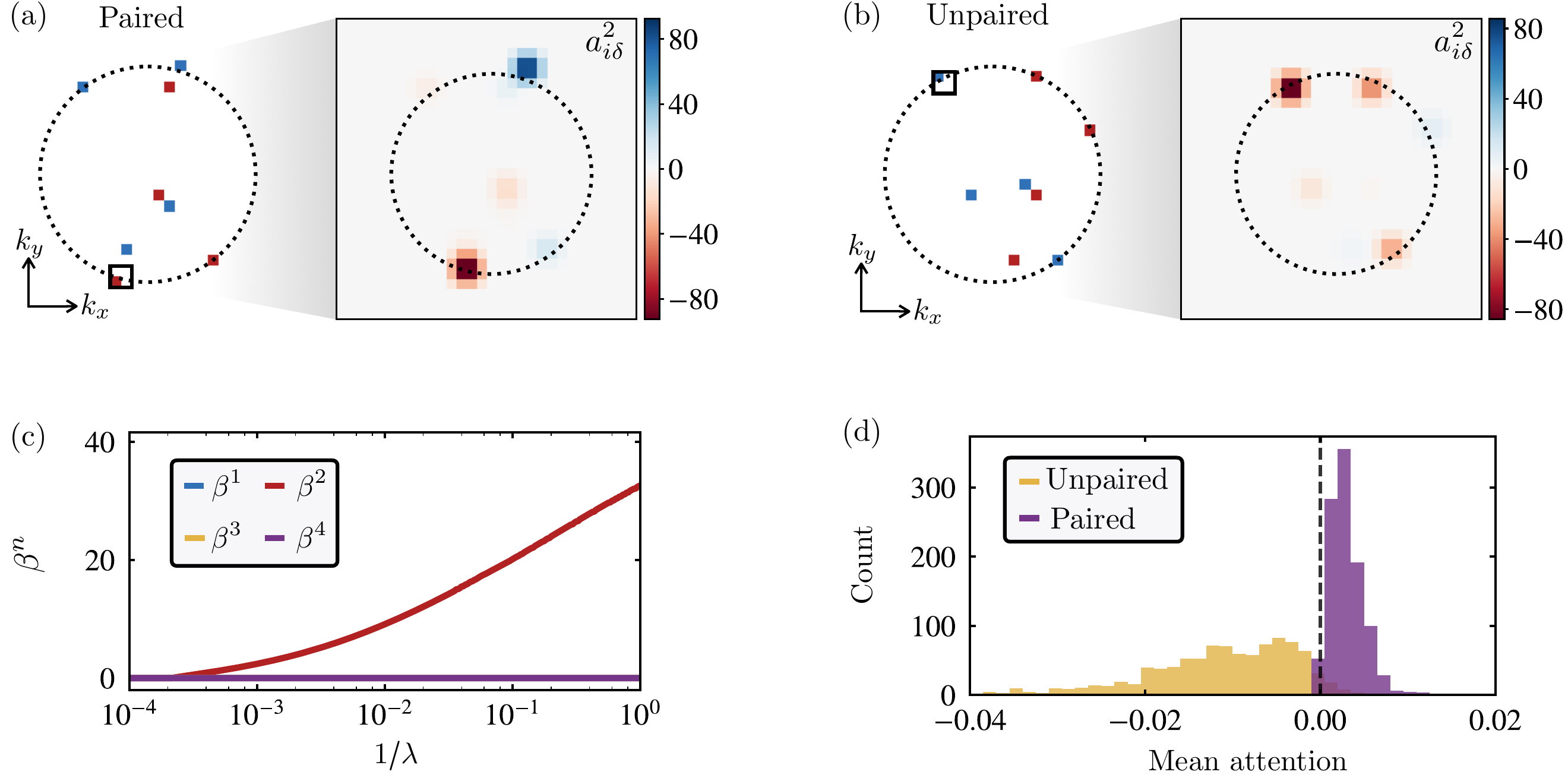}
    \caption{\textbf{Detecting Cooper pairs in momentum space.} (a) Left: a simulated snapshot of a Fermi gas in momentum space with two Cooper pairs on the Fermi surface, mimicking the data obtained in~\cite{Holten_2022}. Blue (red) pixels denote electrons with spin up (down). The Fermi surface is highlighted by the black dashed line. Right: Attention map with respect to the black square (whose patch label is given by $\delta$). The CoTra is seen to attend to the spin on the other side of the Fermi surface, with the opposite attention weight sign. (b) The same for an unpaired snapshot. Here, though the attention mostly focuses on the Fermi surface as in the paired state, all attention weights are negative. (c) Regularization path analysis indicating $2^{\text{nd}}$ order correlations to be crucial for classification. (d) Mean of the $2^{\text{nd}}$ order attention map for the two data sets; while in the paired state the average attention is sharply peaked around slightly positive values, in the unpaired state average values are broadly distributed and negative. A vanishing mean separates the two data sets, underlined by the black dashed line.}
    \label{fig:cp_results}
\end{figure*}

In order to check which particular four-point correlator the model learns, we construct the fourth-order attention weights $a^4 = \frac{1}{\sqrt{d}}Q^{1}K^{3^T}$, given by (see Appendix~\ref{sec:C4})
\begin{equation}
\begin{aligned}
    C^{^{(4)}}_{hers} = \bigg|\frac{1}{M} \sum_{i=1}^M&\sum_{kln=1}^{d} \sum_{abcm=1}^{d}\sum_{pqto=1}^{d}W^P_{hp}W^{P^T}_{ae}W^{P}_{rk} \\ &  W^{P^T}_{ls} W^{1Q}_{km}W^{2Q^T}_{ca}
    W^{3Q}_{pt}W^{1K^T}_{ml}W^{2K}_{bc}\\ &W^{3K^T}_{qt}
    W^{1V}_{nb}W^{2V^T}_{qo}(\text{PE})_{in}(\text{PE})^T_{oi}\bigg|,
    \label{eq:weights4}
\end{aligned}
\end{equation}
where the indices $h,e,r,s$ run through the $8$ elements inside a single patch, and we take an average over all the patches to compute $C^{^{(4)}}_{hers}$. Note that the only dependency on patches comes from absolute positional embedding in this case. We show the intra-patch weights as given in Eq.~\eqref{eq:weights4} in Fig.~\ref{fig:ilgt_results}~(c), and find that the model is giving the most significant weight to the correlator corresponding to the term $(h,e,r,s) = (2,5,6,7)$, which matches Gauss' law [cf. the labeling in each unit cell as shown in the inset of Fig.~\ref{fig:ilgt_results}~(c)].

We note that for the case of the $\mathbb{Z}_2$ Ising LGT, we additionaly use a $Layer~Norm$ after the query and key projections, which we identify as crucial for a successful classification (without it, the network fails to identify the gauge structure, leading to an accuracy of $50\%$). We further find that the network's output depends on the choice of hyperparameters: runs with various choices of parameters filter out $2^{\text{nd}}$, $3^{\text{rd}}$ as well as $4^{\text{th}}$ as the most important correlations. Nevertheless, in Appendix~\ref{sec:HD}, we show that for all cases, corresponding results are interpretable and that, indeed, classification can be achieved in various ways by looking at different orders of correlations for the classical $\mathbb{Z}_2$ LGT.    

\section{Learning Non-Local Structures}\label{nonlocal_obs}
So far, we have seen that the CoTra can detect local correlations as well as local gauge structures through the evaluation of local $n$-point correlations. In this section, we consider datasets containing snapshots that can be distinguished only by detecting non-local structures. This demonstrates the ability of the Transformer to capture long-range dependencies, which constitutes a challenge for many architectures and thus establishes the CoTra as a versatile analysis tool for quantum matter data.

\subsection{Cooper Pairs}
Our first testing case is the detection of non-local structures of Fermi surfaces in BCS-type states that feature bound Cooper pairs~\cite{Bardeen1957}.  
This is in particular motivated by recent experimental observations of Fermi surface correlations in a mesoscopic 2D Fermi gas~\cite{Holten_2022}. There, using a 2D optical dipole trap-optical tweezer setup, momentum-resolved snapshots have been obtained through time of flight (TOF) expansion measurements. In Ref.~\cite{Holten_2022}, single particle momentum correlations are analyzed in the experimental data. At weak attractions, the particles show no correlations; however, at sufficiently strong interaction strengths, they form opposite momentum opposite spin pairs across the Fermi surface. Here, we generate synthetic TOF data akin to the experimental results in Ref.~\cite{Holten_2022} and train the CoTra to analyze whether it can successfully capture these long-range correlations in continuous image-like data.

To this end, we again generate two classes of data: In both cases, the momentum space snapshots are generated by distributing $+1$ for spin-up and $-1$ for spin-down fermions on a $30\times30$ grid with a Fermi momentum of $10$ units. The paired state features two Cooper pairs on the Fermi surface and four particles randomly
distributed inside the Fermi surface. Images in the second class also have two spin-up and two spin-down fermions on the Fermi surface, but none of these form a pair. Exemplary snapshots for paired and unpaired states are shown in the left-hand side of Fig.~\ref{fig:cp_results}~(a) and (b), respectively.

The hyperparameters for this data are $d=16, \text{patch size} =2, \lambda=10^{-4}, \text{ and } lr=10^{-2}$; we obtain an accuracy of $98.1\%$. The regularization path analysis shown in Fig.~\ref{fig:cp_results}~(c) indicates that the Transformer is using second-order correlations to make its decision. The distributions of mean second order attention maps (mean over the whole attention map matrix) are shown in Fig.~\ref{fig:cp_results}~(d). Similar to earlier examples, the paired state has a sharp distribution of attention values, which is largely separated from a broad distribution of the unpaired state.

We obtain the attention map with respect to a given patch $\mathbf{\delta}$ through 
\begin{equation}
\begin{aligned}
    a^2_{i\mathbf{\delta}} = \sum_{rs=1}^{cp^2}\sum_{klsm=1}^{d}&x_{ir}x^T_{s\mathbf{\delta}}P_{ik}P^T_{l\mathbf{\delta}} \\ &W^p_{rk}W^{p^T}_{ls}W^{1q}_{km}W^{1k^T}_{ml}.
    \label{eqn: attnwpatch}
\end{aligned}
\end{equation}
The index $i$ runs through all the patches, thereby giving us correlations between patch $\mathbf{\delta}$ and all others. Thus, by focusing on quantities like Eq.~\eqref{eqn: attnwpatch}, long-range correlations can be analyzed. The right-hand side of Fig.~\ref{fig:cp_results}~(a) and (b) show the resulting second-order attention maps.

\begin{figure*}
    \centering
\includegraphics[width=0.9\textwidth]{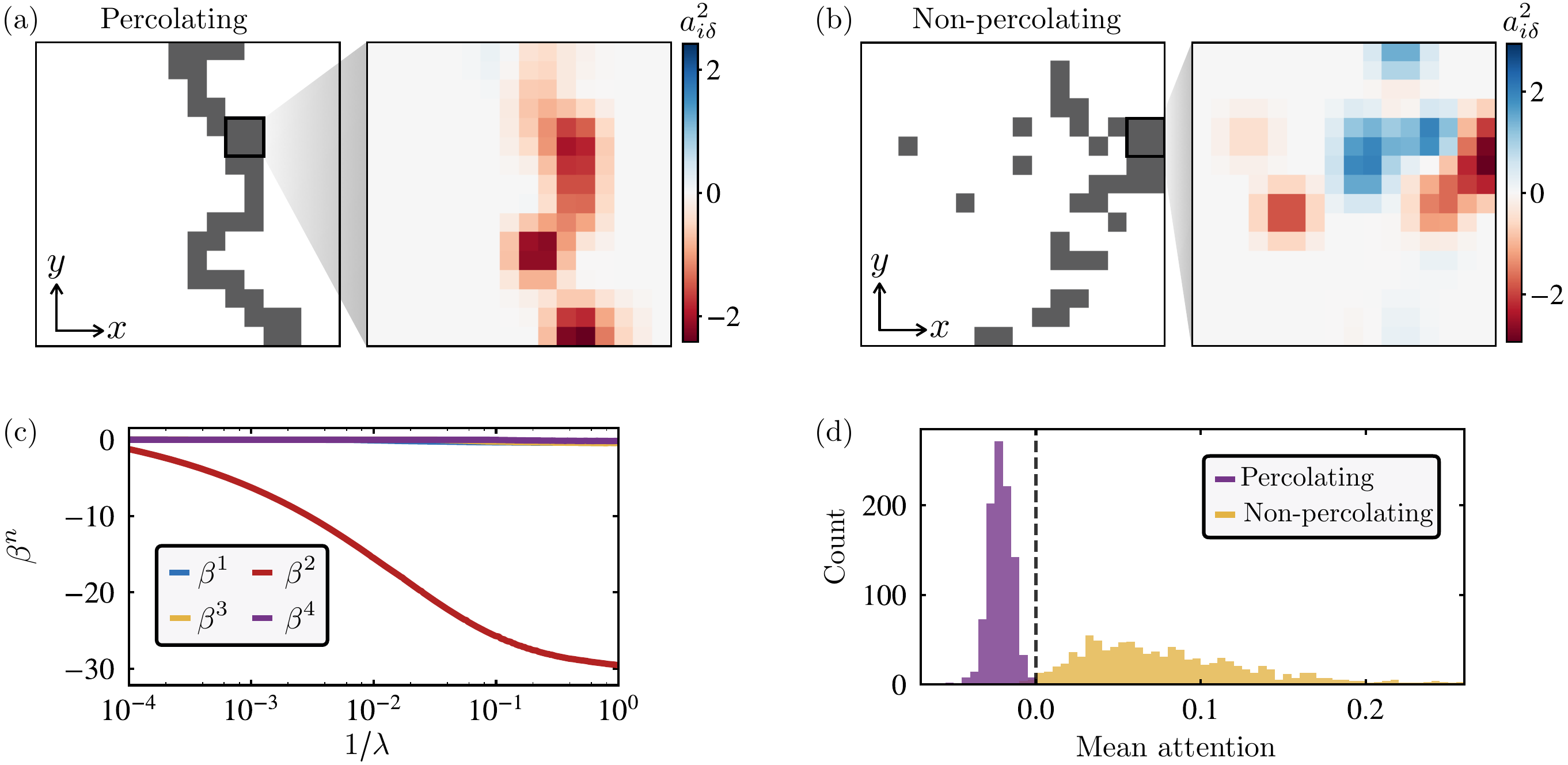}
    \caption{\textbf{Detecting percolation.} (a) Left: percolated structure of a binary snapshot. Right: Attention map $a_{\delta i}$ with respect to the black square (whose patch label is given by $\delta$). The CoTra is seen to attend to the percolating structure, with uniform attention weight sign. (b) The same for a non-percolated snapshot. Here, the attention is focused mostly around the reference patch $\delta$, with a seemingly random sign structure. (c) Regularization path analysis indicating $2^{\text{nd}}$ order correlations to be crucial for classification. (d) Mean of the $2^{\text{nd}}$ order attention map for the two data sets; while in the percolated phase the average attention is sharply peaked around slightly negative values, in the non-percolated state average values are broadly distributed and positive. A vanishing mean separates the two data sets, underlined by the black dashed line.}
    \label{fig:perc_results}
\end{figure*}

For the paired snapshot, the attention is clearly given to the Cooper pair; see Fig.~\ref{fig:cp_results}~(a). In contrast, no such structure exists in the unpaired state, see Fig.~\ref{fig:cp_results}~(b). A particular feature of the attention maps is their sign structure: In the paired state, attention weights of the Cooper pair are of opposite sign, while in the random state, all weights are negative. This, in turn, leads to a separation of the mean attention values, and hence the separation of the corresponding histograms, cf. Fig.~\ref{fig:cp_results}~(d). We note that the above holds independent of the reference patch $\delta$. In Appendix~\ref{sec:CP_app}, we show attention maps when using a reference patch containing a particle inside the Fermi surface. In this case, the attention is on both Cooper pairs, while no particular structures exist for the unpaired state. 

\subsection{Percolation}
Percolation theory, first introduced by
Flory and Stockmayer in the 1940s~\cite{Flory1941}, considers the formation of clusters that eventually span an extended system in a percolating cluster~\cite{Hu1984}. Naturally, the distinction of whether a particular snapshot is percolating or not constitutes a problem that requires the evaluation of non-local correlations. In this section, we train the CoTra with synthetic data of percolating and non-percolating binary images on a 2D lattice.

In particular, we focus on a simplified percolating system with only one percolating path (directed percolation). We generate a $16\times16$ lattice of random spins directed from one edge to the opposite edge (i.e. we generate random percolating structures). Half of the generated percolating images are then randomly shuffled such that they become non-percolating. The left-hand sides of Fig.~\ref{fig:perc_results}~(a) and (b) show exemplary percolating (non-percolating) snapshots. We note that due to our construction of the snapshots, the total number of grey pixels is the same; this prevents the network from simply counting the number of active pixels.

We train the CoTra with hyperparameters $d=16, \text{patch size} =2, \lambda=10^{-3}, \text{ and }lr=10^{-2}$. We obtain an accuracy of $99.3\%$. After training, a regularization path analysis shown in Fig.~\ref{fig:perc_results}~(a) indicates second-order correlations to be crucial for the classification. Mean second-order attention values are seen to separate, whereby percolating (non-percolating) images have a sharp (broad) distribution, see Fig.~\ref{fig:perc_results}~(b). We emphasize the generality of this observation and suspect that a global underlying structure typically leads to distributions of the CoTra mean attention map values that have small widths.  

We plot second-order attention maps given by Eq.~\eqref{eqn: attnwpatch} on the corresponding right-hand sides of Fig.~\ref{fig:perc_results}~(a). The reference patch $\delta$ is given by the black square in the snapshot images. In the percolating case, the attention mechanism indeed puts emphasis on the percolating structure with uniform sign. In case of the non-percolating image, the attention seems to be localized relatively close to the reference patch, with a seemingly random sign structure. We note that, by utilizing the capability of analyzing the dependencies between all patches, second-order inter-patch (i.e., long-range) correlations allow the network to make a classification. We note that when choosing a reference patch that does not contain any pixel of the percolating cluster, the attention maps is zero. 

For both Cooper pairs and percolation, the sign of the attention map is seen to be an important factor contributing to the classification. In both cases, we see that the classification boundary lies around zero [see the black dashed lines in Fig.~\ref{fig:cp_results}~(d) and Fig.~\ref{fig:perc_results}~(d)]. By assigning a certain sign structure to the inter-patch attention, the CoTra is able to reliably detect non-local structures through inter-patch correlations.  

\section{Summary and Outlook}\label{outlook}
We have introduced a Transformer-inspired machine learning framework to analyze image-like quantum matter data with a tailored attention mechanism that yields full interpretability in terms of correlation functions. Combined with a regularization path analysis, we have shown that the network is able to identify both local and non-local structures in a variety of testing cases. We have demonstrated that the network can learn local correlations, such as present in the AFM Heisenberg model, and that it can identify gauge constraints, e.g. in 1D and 2D lattice gauge theories. We further demonstrated the network's capabilities of detecting non-local dependencies and correlations, by training on images of correlated Fermi surfaces featuring BCS pairs as well as percolating and non-percolating images. 

Our proposed architecture can readily be used for experimental image-like data, for instance, real- and momentum-space snapshots of ultracold atoms~\cite{Fu_2020, lunt2024} or scanning tunneling microscopes. This may help to reveal hidden order, topological phases, and other global, non-local structures such as symmetries of e.g. charge density wave phases~\cite{Fujita2014}. The CoTra may further be used for learning and interpreting disentangling gates, akin to the reinforcement learning protocol demonstrated in~\cite{tashev2024}. Recently, it has been proposed that percolation can be used as an effective order parameter for confinement in $\mathbb{Z}_2$ LGTs~\cite{Linsel2024}. It would be interesting to see how the CoTra performs when given finite temperature snapshots of confined and deconfined phases, and if any connection to percolation can be made from the resulting attention maps. It would further be interesting to combine the CoTra with a confusion learning scheme akin to~\cite{Henning2023}, which allows for a fully unsupervised phase detection and classification. This, in particular, can be useful to analyze doping and temperature scans of the strongly correlated Fermi-Hubbard model, where non-local, hidden structures are believed to play an essential role in the low-doping regime~\cite{Hilker2017, Schloemer2022_recon}. \\ \\

\textbf{Acknowledgements.--}
We thank Simon M. Linsel for valuable comments. This research was funded by the Deutsche
Forschungsgemeinschaft (DFG, German Research Foundation) under Germany’s Excellence Strategy – EXC2111 – 390814868, as well as the European Research Council (ERC) under the European Union’s Horizon 2020 research and innovation programme (grant agreement number 948141 — ERC Starting Grant SimUcQuam). AS acknowledges support by the Graduiertenkolleg of the University of Regensburg (IRTG SFB 1277), as well as the KVPY program of DST (Department of Science and Technology) Govt of India.  BH acknowledges support by Excellence Cluster ORIGINS, funded by the Deutsche Forschungsgemeinschaft (DFG, German Research Foundation) under Germany’s Excellence Strategy - EXC-2094-390783311.

\appendix
\section{Correlator Weights}
We here derive the expressions for the attention maps as well as correlator weights presented in the main text, see i.p. Eqs.~\eqref{eq:C2}, \eqref{eq:weights3}, \eqref{eq:weights4}.
\subsection{$2^{\text{nd}}$ Order Correlators}
\label{sec:C2}

Let $x$ be the input image in $\mathbb{R}^{(h,w,c)}$, where $h,w,c$ represents the height, width and number of channels. First, we patch this image with a patch size of $p$; the flattened image will then be of size $(\frac{hw}{p^2},cp^2)$ for a 2D snapshot and $(h/p,cp)$ for a 1D snapshot. We denote the number of patches ($\frac{hw}{p^2}$) with $M$. 

Now we linearly project these flattened patches to a hidden dimension $d$ and multiply with a learnable position encoding $P$ in an element-wise fashion,
\begin{align}
    X^p &= P*W^pX \\ \nonumber
    X^p_{ij} &= \sum_r^{cp^2} x_{ir}W^p_{rj}P_{ij}.
\end{align}
Here, $W^p$ is the projection matrix in $\mathbb{R}^{(cp^2,d)}$ and $X$ denotes the flattened patches. The learnable position encoding helps in more accurate classification since it encodes each patch with a certain weight.

After this, we introduce three independent linear projections by $W^{\alpha Q}, W^{\alpha K}, W^{\alpha V}$ to form the query, key and value, as in standard Transformer architectures~\cite{vaswani2023attention}. $\alpha$ denotes the particular Transformer layer that the projection is applied to.

\begin{align}
    Q^1_{ij} &= \sum_k^d\sum_r^{cp^2} x_{ir}W^p_{rk}P_{ik}W^{1Q}_{kj} \\
    K^1_{ij} &=  \sum_l^d\sum_s^{cp^2} x_{is}W^p_{sl}P_{il}W^{1Q}_{lj}\\
    V^1_{ij} &=  \sum_n^d PE_{in}W^{1V}_{nj}
\end{align}
Where the value ($V^1$) is derived from the sine-cosine positional encoding, see Eq.~\eqref{eq:sincos}. Now, we calculate the $2^{\text{nd}}$ order attention map $a^2$ (setting the constants ($d$) to 1 for simplicity) by
\begin{align}
    a^2 =& Q^1K^{1^T} \nonumber \\
    a^2_{ij}  = &\sum_{rs=1}^{cp^2}\sum_{klsm=1}^{d}x_{ir}x^T_{sj}P_{ik}P^T_{lj}W^p_{rk} \nonumber\\ &W^{p^T}_{ls}W^{1q}_{km}W^{1k^T}_{ml}
    \label{eq:a2A}
\end{align}
The $2^{nd}$ order weight matrix for a given patch can be constructed from the coefficients of $a^2$ by fixing $i=j$ and then taking an average over all $M$ patches $i$. In the analysis in the main text, we focus on the absolute value of all the $n^{th}$ order correlator weights $C^{^n}$.

\begin{align}
    C^{^{(2)}}_{rs} =\left|\frac{1}{M} \sum_{i=1}^{M} \sum_{klm=1}^{d} P_{ik}P_{li}^TW^P_{rk}W^{P^T}_{ls}W^{1Q}_{km}W^{1K^T}_{ml}\right|,
    \label{eq:C2Ap}
\end{align}

The output of the first layer $X^2$ can be written as
\begin{align}
    X^2 = &a^2V^1 \nonumber \\
    X^2_{ij} = &\sum_{g=1}^{M}\sum_{rs=1}^{cp^2}\sum_{klsmn=1}^{d} x_{ir}x^T_{sg}P_{ik}P^T_{lg}W^p_{rk}\nonumber\\ &W^{p^T}_{ls}W^{1q}_{km}W^{1k^T}_{ml}(PE)_{gn}W^{1v}_{nj}
    \label{x2}
\end{align}

\subsection{$3^{rd}$ Order Correlators}
\label{sec:C3}
The key projection in the $2^{\text{nd}}$ layer is derived from $X^2$ while the query projection remains the same, ie from $X^1$ itself but with a different projection matrix $W^{2Q}$,
\begin{align}
    K^{2}=& X^2W^{2k} \nonumber\\ 
    K^{2}_{ij} = &\sum_{g=1}^{M}\sum_{rs=1}^{cp^2}\sum_{klsm=1}^{d} \sum_{nb=1}^{d} x_{ir}x^T_{sg}P_{ik}P^T_{lg}W^{p}_{rk}W^{p^T}_{sl}\nonumber \\ 
    &W^{{1q}}_{km}W^{1k^T}_{ml}(PE)_{gn}W^{{1v}}_{nb}W^{2k}_{bj}.
\end{align}

The $3^{rd}$ order attention map can be constructed by

\begin{align}
    a^3  =& Q^2K^{2^T}  \nonumber\\
    a^3_{ij} =& \sum_{g=1}^{M}\sum_{ers=1}^{cp^2}\sum_{klsn=1}^{d} \sum_{abcm=1}^{d}x_{ie}x^T_{rj}x_{ng}P_{ia}P^T_{kj}P_{gl}W^p_{ea} \nonumber\\&W^{p^T}_{kr}W^{p}_{sl} 
    W^{1q^T}_{mk}W^{2q}_{ac}W^{1k}_{lm}W^{2k^T}_{cb}W^{1v^T}_{bn}(PE)^T_{ng},
\end{align}
where $Q^2$ is the $2^{nd}$ layer query projection of $X^1$.
The $3^{rd}$ order correlator weights for a given patch can be constructed from learnable coefficients of the $3rd$ order attention map by fixing $i=j=g$ and then taking a mean over all the $M$ patches.
\begin{equation}
\begin{aligned}
        C^{^{(3)}}_{ers} = &\bigg|\frac{1}{M} \sum_{i=1}^M\sum_{kln=1}^{d} \sum_{abcm=1}^{d}P_{ia}P^T_{ki}P_{il}W^P_{ea}W^{P^T}_{kr}\\ &W^P_{sl}W^{1Q^T}_{mk}W^{2Q}_{ac}W^{1K}_{lm}W^{2K^T}_{cb}W^{1V^T}_{bn}(\text{PE})_{ni}\bigg|,
\end{aligned}
\end{equation}
The output of the $2^{\text{nd}}$ layer $X^3$ can be written as 

\begin{align}
     X^3_{ij}=&\sum_{gf=1}^{M}\sum_{ers=1}^{cp^2}\sum_{klsn=1}^{d} \sum_{abcmo=1}^{d}x_{ie}x^T_{rf}x_{gs}P_{ia}P^T_{kf}\\ \nonumber&P_{gl}W^p_{ea}W^{p^T}_{kr} W^{p}_{sl}W^{1q^T}_{mk}
     W^{2q}_{ac}W^{1k}_{lm}W^{2k^T}_{cb}W^{1v^T}_{bn}\\ \nonumber&(PE)^T_{ng}W^{2v}_{oj}(PE)_{fo}.
\end{align}
\subsection{$4^{\text{th}}$ Order Correlators}
\label{sec:C4}
Similar to constructing the previous order correlator weights, we can obtain the key projection from $X^3$ while keeping the query projection the same. Note that we do not use any learnable positional encoding for the Ising LGT.

\begin{align}
    K^3 =& X^3W^{3k} \\ \nonumber
    K^{3}_{ij} =&\sum_{gf=1}^{M}\sum_{ers=1}^{cp^2}\sum_{klsn=1}^{d} \sum_{abcm=1}^{d}\sum_{oq=1}^{d}x_{ie}x^T_{rf}x_{ns}\\ \nonumber&W^p_{ea} W^{p^T}_{kr}W^{p}_{sl}W^{1q^T}_{mk}W^{2q}_{ac}W^{1k}_{lm}
     W^{2k^T}_{cb}W^{1v^T}_{bn}\\ \nonumber&(PE)^T_{nf}W^{2v}_{oq}(PE)_{fo}W^{3k}_{qj}
\end{align}
The $4^{\text{th}}$ order attention map is written as
\begin{align}
    a^4 =& Q^3K^{3^T} \\ \nonumber
    a^4_{ij} =& \sum_{gf=1}^{M}\sum_{ers=1}^{cp^2}\sum_{klsn=1}^{d} \sum_{abcm=1}^{d}\sum_{pqto=1}^{d}x_{ih}x^T_{ej}x_{qr}x^T_{sn}\\ \nonumber
   &W^p_{hg}W^{p^T}_{ae} W^{p}_{rk}W^{p^T}_{ls}W^{1q}_{km}W^{2q^T}_{ca} W^{3q}_{gp}W^{1K
   k^T}_{ml}\\ \nonumber
   &W^{2k}_{bc}W^{3k^T}_{pq}W^{1v}_{nb}W^{2v^T}_{qo}(PE)_{gn}(PE)^T_{of}.
\end{align}
Where $Q^3$ is the $3^{rd}$ layer query projection of $X^1$.
The $4^{\text{th}}$ order correlator weights can then be written as

\begin{equation}
\begin{aligned}
    C^{^{(4)}}_{hers} = \bigg|\frac{1}{M} \sum_{i=1}^M&\sum_{kln=1}^{d} \sum_{abcm=1}^{d}\sum_{pqto=1}^{d}W^p_{hp}W^{p^T}_{ae}W^{p}_{rk} \\ &  W^{p^T}_{ls} W^{1q}_{km}W^{2q^T}_{ca}
    W^{3q}_{pt}W^{1k^T}_{ml}W^{2k}_{bc}\\ &W^{3k^T}_{qt}
    W^{1v}_{nb}W^{2v^T}_{qo}(\text{PE})_{in}(\text{PE})^T_{oi}\bigg|.
    \label{eq:weights4A}
\end{aligned}
\end{equation}
\section{Discussion on hyperparameter dependence}
\label{sec:HD}
As noted in the main text, we find that especially for the Ising $\mathbb{Z}_2$ LGT, the results depend on the choice of hyperparameters. One possible explanation for this is our use of the $Layer Norm$, which is crucial for a successful classification in the case of the Ising LGT. The use of $Layer Norm$ complicates the interpretability of the network because we do not account for the weights and bias of the normalization in our analysis. By varying the hyperparameters, we indeed observe that in different runs the Transformer focuses on $2^{\text{nd}}$, $3^{\text{rd}}$ and $4^{\text{th}}$ order correlators to classify the Ising LGT. In the following, we look in more detail into the output of the Transformer for all of the above cases and show that the network's decisions remain interpretable.

\begin{figure}
    \centering
    \includegraphics[width=8.8cm]{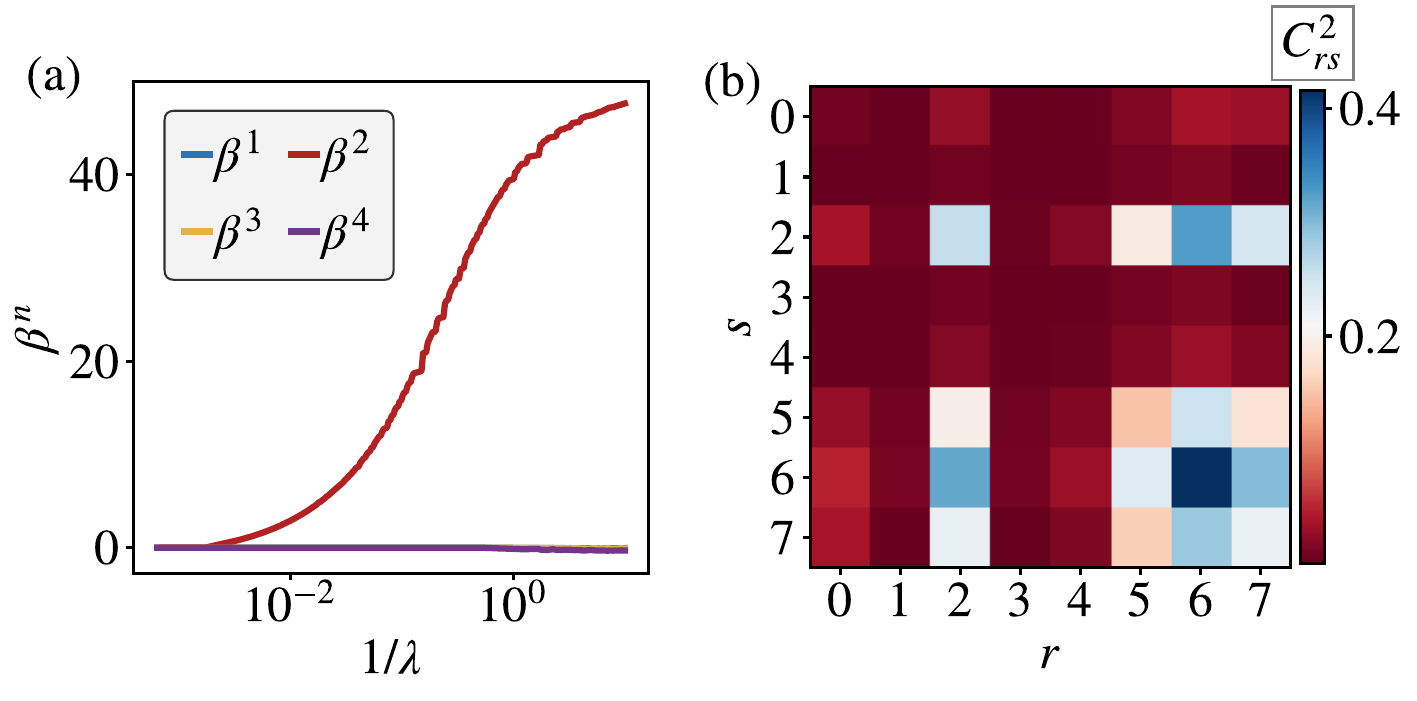}
    \caption{a) Regularization path analysis indicating $2^{nd}$ order terms to be important b) $2^{nd}$ order weights from the attention map}
    \label{fig:igt_2}
\end{figure}

\begin{figure}
    \centering
    \includegraphics[trim=1cm 1cm 1cm 1cm,width=8.7cm]{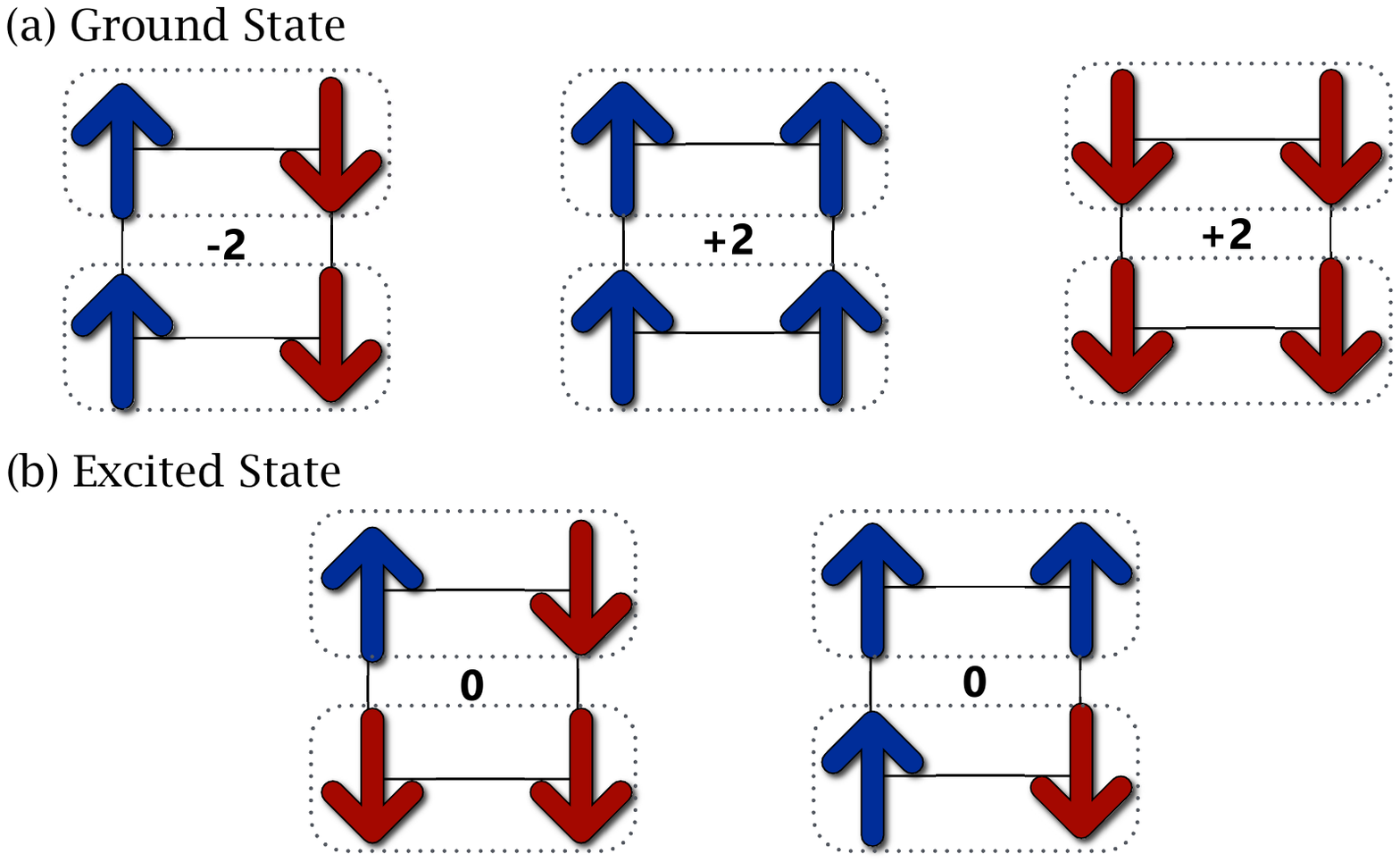}
    \caption{a) Schematic showing possible spin alignments on a plaquette (blue indicates spin up and red indicates spin down) for the ground state; the resulting sums of the 2-point terms are shown in the centres. b) Additional possible configurations in an excited state, which does not fulfil the gauge constraint on each plaquette.}
    \label{fig:igt_expl}
\end{figure}

Fig.~\ref{fig:igt_2} shows the results for the hyperparameters:  $d=4, \text{patch size} =2,  \lambda=1e^{-3}, lr=0.005$. For this set of hyperparameters, the network uses $2^{\text{nd}}$ order correlations to discriminate the phases. Though this may seem surprising, it is indeed possible to discriminate the two phases by adding certain $2$-point correlations: If we look at the sum of the (horizontal) product of two links - call it $g^2$ - on the plaquette, we get the following criteria (see Fig.~\ref{fig:igt_expl}):  
\begin{align}\label{Eqn:g2}
    g^2 &= \pm2 \text{~~for states that globally fulfil GL}\\ \nonumber 
    g^2 &= 0,\pm2 \text{~~for excited states}.
\end{align}
\begin{figure}
    \centering
    \includegraphics[width=8cm]{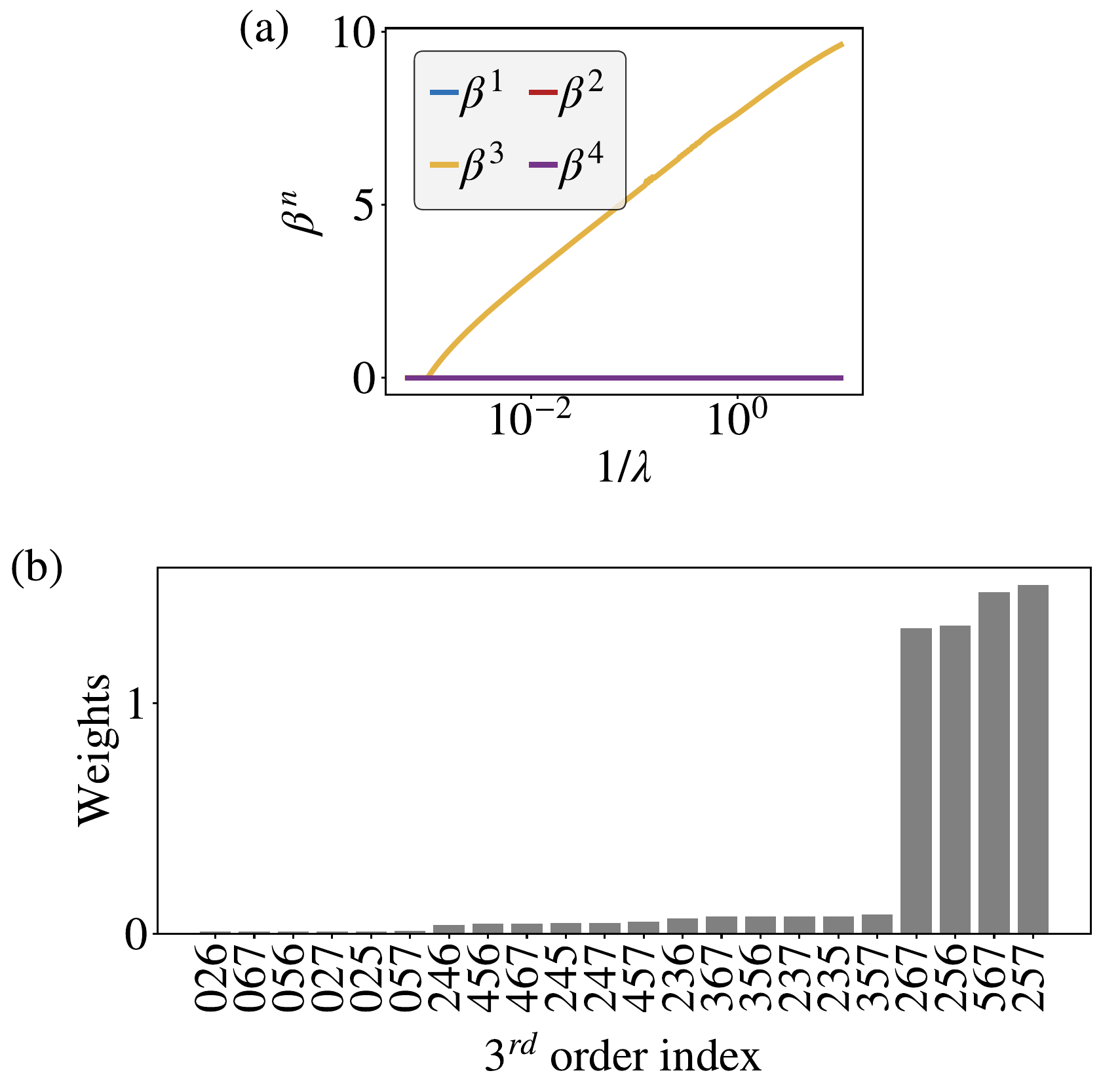}
    \caption{(a) Regularization path analysis indicating $3^{\text{rd}}$ order term to be important. (b) $3^{rd}$ order weights from the attention map (only values above a threshold of $0.005$ are plotted). The network calculates 3-point correlators on the plaquette, from which it can judge whether the gauge constraint is fulfilled or not.}
    \label{fig:igt_3}
\end{figure}

Indeed, this is similar to Gauss' law Eq.~\eqref{Gauss's_law}. In the weight matrix $C^2_{rs}$, shown in Fig.~\ref{fig:igt_expl}~(b), we see that 2-point correlations involving the plaquette sites $2,5,6,7$ are used. By adding these correlations, the Transformer is able to reliably detect the local gauge constraints.

Fig.~\ref{fig:igt_3} shows the results for the hyperparameters  $d=28, \text{patch size} =2, \lambda=1e^{-4}, lr=0.01$; here, the network utilizes 3-point correlations for classification.

The reasoning for the $2^{\text{nd}}$ correlations can also be extended to the $3^{\text{rd}}$ order. The sum of $3$-point terms $g^3$ on a plaquette can be categorized as
\begin{align}\label{Eqn:g3}
    g^3 &= 0,\pm4 \text{~~for ground states}\\ \nonumber 
    g^3 &= \pm2 \text{~~for excited states}.
\end{align}
As we can see from the correlation weights for the $3^{rd}$ order terms, Fig.~\ref{fig:igt_3}~(b), the larger weights are associated with the terms $256$, $257$, $267$ and $567$. Thus, by summing over these 3-point correlations, perfect accuracy can be achieved.  
\section{Varying reference patch for Cooper pair data}
\label{sec:CP_app}
\begin{figure}
    \centering
    \includegraphics[width=0.8\columnwidth]{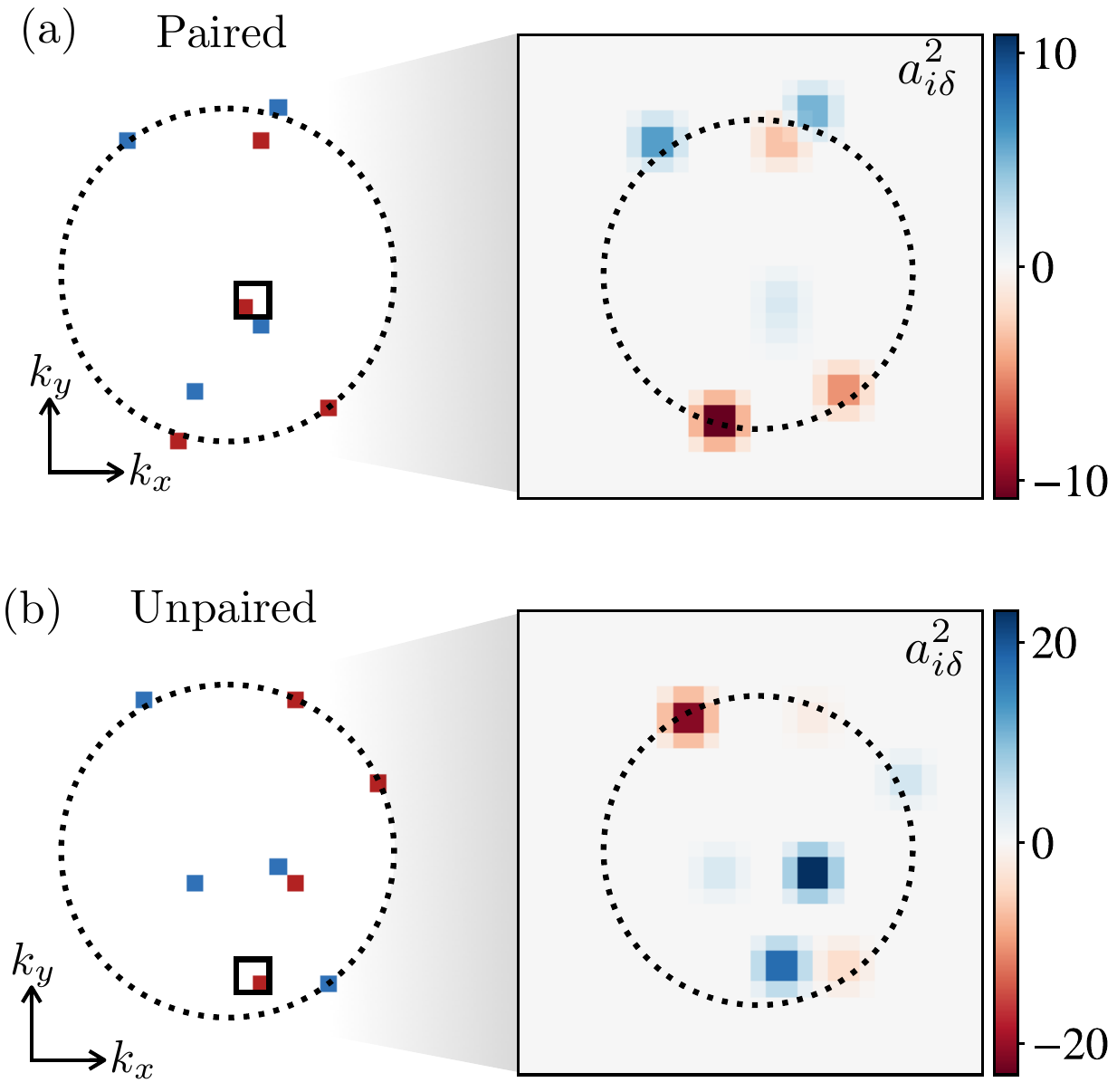}
    \caption{(a) Attention map for Cooper paired state w.r.t the patch denoted by the black square. Attention is seen to be put on the two Cooper pairs on the Fermi surface, regardless of the reference patch. (b) Attention map for the unpaired state w.r.t the patch denoted by the black square. No structures as observed in (a) can be seen.}
    \label{fig:CP_attn_i}
\end{figure}

\begin{figure*}
    \centering
    \includegraphics[trim=1cm 5cm 1cm 3cm,width=17cm]{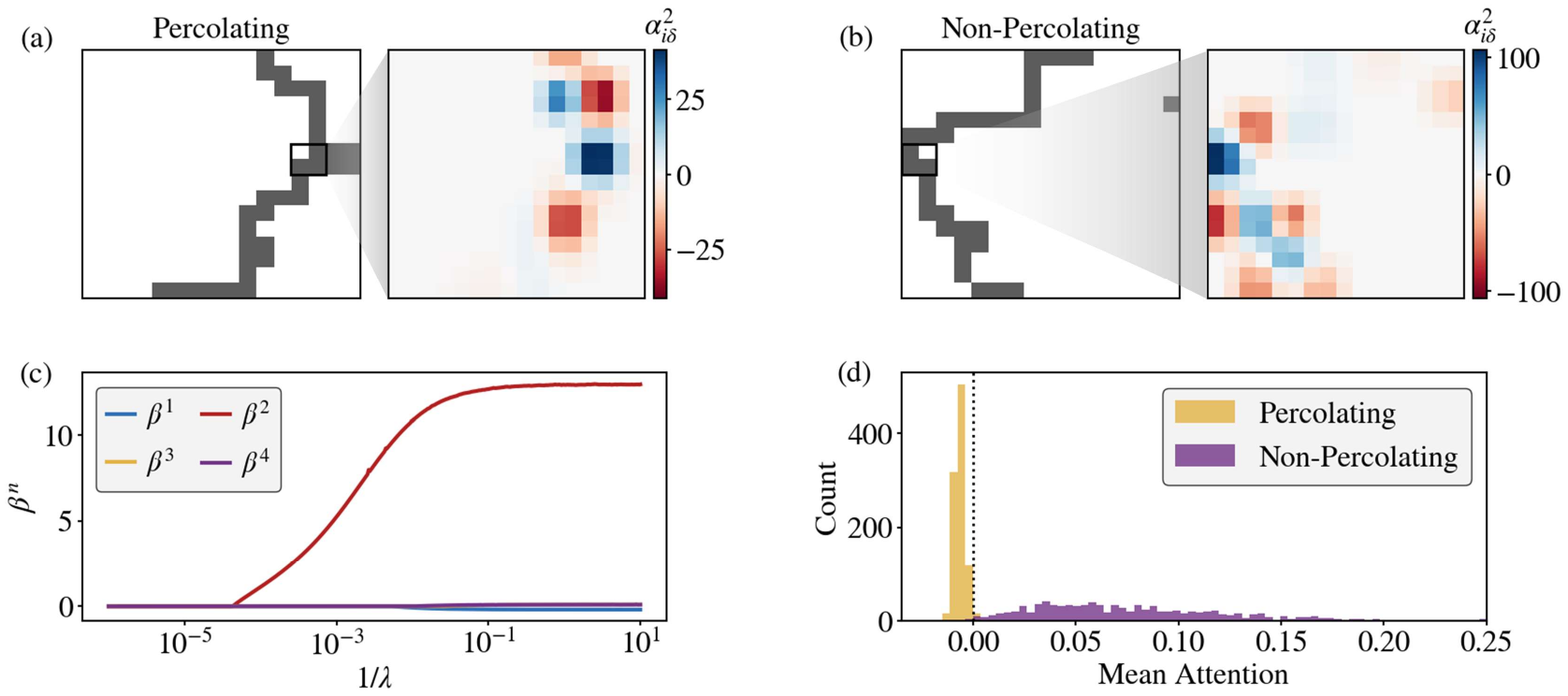}
    \caption{a) Attention map w.r.t the highlighted patch for the percolating state, b) attention map w.r.t the highlighted patch for the non-percolating state, c) regularization path indicating $2^{nd}$ order correlators to be important, d) Mean of the $2^{nd}$ order attention map for the two data sets}
    \label{fig:perc_attn_sup}
\end{figure*}
Fig.~\ref{fig:CP_attn_i}
shows the attention map w.r.t a particle inside the Fermi surface. For a snapshot that features Cooper pairs, the attention is - regardless of the reference patch - seen to focus on the paired electrons on the Fermi surface, see Fig.~\ref{fig:CP_attn_i}~(a). On the other hand, for an unpaired snapshot, Fig.~\ref{fig:CP_attn_i}~(b), no such structures are observable.

\section{Additional Results}
\subsection{Percolation}
In this section, we train the model with slightly modified percolation data in which the percolating path is broken at very few points. We get similar results to the main text data with $2^{nd}$ order correlations to be the most important.

The hyperparameters used are $d=16, \text{patch size} =2, \lambda=2\times10^{-3}, \text{ and } lr=5\times10^{-3}$; we obtain an accuracy of $95.78\%$. This indicates that the CoTra is able to precisely classify percolation in this dataset.
Fig.~\ref{fig:perc_attn_sup} (a,b) shows $2^{nd}$ attention maps for the percolating and non-percolating states. While Fig.~\ref{fig:perc_attn_sup} (c,d) shows the regularization path and histogram of mean attention.

\subsection{Staggered Magnetization}
In this section, we train the model with snapshots of $16\times16$ lattice having staggered magnetization equal to zero against snapshots with staggered magnetization equal to $0.015625$. 

The hyperparameters used are $d=4, \text{patch size} =2, \lambda=10^{-4}, \text{ and } lr=10^{-1}$. The model is able to successfully classify the dataset with $99.38\%$ accuracy. 
\begin{figure}
    \centering
    \includegraphics[width=8.5cm]{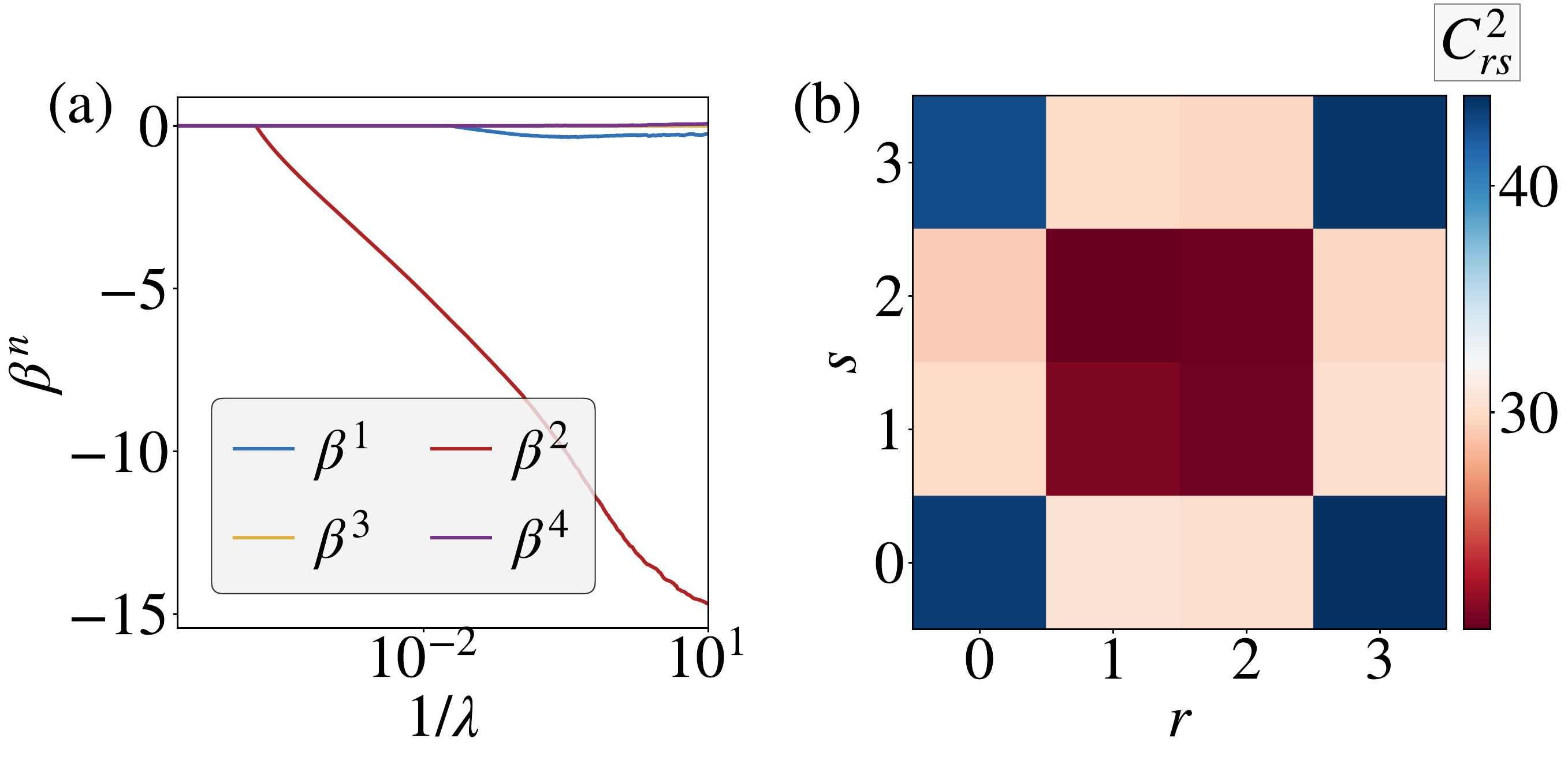}
    \caption{a) \textbf{Detecting staggered magnetization.} Regularization path analysis b) Corresponding $2^{nd}$ order correlation weights. }
    \label{fig:stg}
\end{figure}
Fig.~\ref{fig:stg} (a) shows the $2^{nd}$ order regularization path, and Fig.~\ref{fig:stg} (b) shows the corresponding order weights. Though we get more than $99\%$ accuracy, the regularization paths and the weights depend on the choice of hyperparameters (we show the results for the lowest hidden dimension).
\section*{References}
\bibliography{library.bib}

\end{document}